\documentclass[final,5p,times,twocolumn]{elsarticle}

\usepackage{amsmath,amssymb}
\usepackage{graphicx}
\usepackage{comment}

\begin{document}

%\begin{frontmatter}

\title{Strained Layer Crystalline Undulator}

\author{A.~Kostyuk}

\ead{andriy.p.kostyuk@gmail.com}

%\noaffiliation

%\affiliation{Autogenstra{\ss}e 11, 65933 Frankfurt am Main, Germany}
\address{Ostmarkstra{\ss}e 7, 93176 Beratzhausen, Germany}

\begin{abstract}
Ultrarelativistic charged particles are predicted to emit hard 
electromagnetic radiation of undulator type while being channeled
in a crystal with periodically bent crystallographic planes.
The recently proposed crystalline undulator with 
the bending amplitude smaller than the distance between
the bent planes and the bending period shorter than
the period of channeling oscillations is far superior to
what was proposed previously. In the same time, it is more 
challenging from the technical point of view because 
its bending period has to be in the sub-micron range.
It is shown that a mixed crystal of silicon-germanium 
with properly varying germanium fraction can have the
necessary bending parameters. Moreover, it is predicted to
be stable against misfit dislocations.
\end{abstract}

\begin{keyword}
strained layer superlattice \sep
crystalline undulator

%\pacs{
%68.65.Ac,% Multilayers
\PACS 
68.65.Cd \sep % Superlattices 
%46.25.-y % Static elasticity
46.25.Cc \sep % Static elasticity: Theoretical studies 
%62.20.D-,% Mechanical properties of solids: Elasticity
61.72.uf \sep % defects in Ge and Si
%61.72.Lk,% Linear defects: dislocations, disclinations 
61.85.+p %,% Channeling phenomena (blocking, energy loss, etc.)
%68.55.Ln,% Thin film structure and morphology. Defects and %
         % impurities: doping, implantation, distribution, concentration, etc. 
%41.60.-m,% Radiation by moving charges
%41.60.Ap,% Synchrotron radiation
%02.70.Uu,% Applications of Monte Carlo methods
%41.75.Fr% Electron and positron beams 
%}
\end{keyword}

\maketitle

Channeling \cite{Lindhard1965,Uggerhoj2005a} of ultrarelativistic 
charged particles through 
a periodically bent crystal can be used 
to generate hard electromagnetic radiation of undulator type
\cite{Kaplin1980a,Baryshevsky1980}. Such a radiation source 
may become a compact and affordable alternative to 
presently existing synchrotron radiation sources that are 
widely used in science and technology \cite{Willmott2011}
but are very big and expensive. Moreover, 
a coherent radiation source, a hard X ray or gamma ray laser,
can be built on the basis of the crystalline undulator \cite{Greiner2011}.

Initially, it was suggested that the charged
projectiles should follow the 
sinusoidal shape of the bent crystallographic
planes of the crystalline undulator
performing nearly harmonic transverse oscillations.
Due to these oscillations, the  spectrum of the
electromagnetic radiation emitted by the particles in the forward direction
was expected to have a narrow peak, similarly as it 
took place in an ordinary (magnetic) undulator 
\cite{Ginzburg1947,Motz1951,Motz1953}.

Several conditions had to be satisfied by the crystal 
and the beam to reach the desirable properties of the spectrum \cite{Korol2004}.
In particular, the bending period had to be much larger than the period of 
channeling oscillations, $\lambda_\mathrm{u} \gg \lambda_\mathrm{c}$. Otherwise
the projectile would not follow the shape of the bent crystal channel.
Additionally, the bending amplitude should be much larger than the 
distance between the guiding crystallographic planes, 
$a_\mathrm{u} \gg d$, to make sure that the spectrum
is dominated by the undulator peak rather than by the channeling radiation,
which is present also in the case of a straight crystal \cite{Kumakhov1976c}.
In the following, such a crystalline 
undulator will be  referred to as LALP CU (\textbf{l}arge \textbf{a}mplitude 
and \textbf{l}ong \textbf{p}eriod \textbf{c}rystalline \textbf{u}ndulator).

A new type of crystalline undulator has been proposed recently 
\cite{Kostyuk2012} (see also \cite{Kostyuk2013d}).
It has been demonstrated by numerical simulations
that a crystalline undulator with a bending period smaller
than the channeling period, 
\begin{equation}
\lambda_\mathrm{u} < \lambda_\mathrm{c}
\label{small_lambdau}
\end{equation}
and the bending amplitude smaller than the channel width
\begin{equation}
a_\mathrm{u} < d
\label{small_au}
\end{equation}
has certain advantages over LALP CU.
In particular, it has a much larger effective number of undulator
periods and it requires much lower beam energy for the production 
of radiation of a given frequency comparing to LALP CU.
The new type of crystalline 
undulators is referred to as SASP CU 
(\textbf{s}mall \textbf{a}mplitude  and \textbf{s}hort 
\textbf{p}eriod \textbf{c}rystalline \textbf{u}ndulator).

Channeling of electrons and positrons in SASP CU
has been simulated with the computer code ChaS (\textbf{Cha}nnelling
\textbf{S}imulator) and the coresponding radiation spectra have been 
calculated \cite{Kostyuk2012}. A comparison to a spectrum of LALP CU has been 
presented in \cite{Kostyuk2013d}.
In contrast to LALP CU, the channeled projectile in the 
SASP CU does not follow 
the shape of the bent planes. Still, the shape of the 
trajectory does contain a Fourier component with the period
equal to that of the undulator bending. As the result, a narrow 
undulator peak is present in the radiation spectrum of SASP CU. 
This result has been confirmed by 
independent computations \cite{Baryshevsky2013a} and 
verified experimentally \cite{Wistisen2014}.
Application of SASP CU to nuclear waste transmutation has
been discussed in \cite{Uggerhoj2015}.

The intensity of the radiation is proportional to the 
fourth power of the frequency.
The frequency of the undulator radiation is higher than
that of the channeling radiation. Therefore, the intensity
of the undulator radiation peak is higher than the channeling one
despite of the smallness of the undulator amplitude with 
respect to the channel width (see \cite{Kostyuk2012} 
and \cite{Kostyuk2013d} for details).

From technological point of view, SASP CU is more challenging 
than LALP CU. For moderate projectile energy
$E \lesssim 1$ GeV, it is necessary to bend the crystal 
with a period 
shorter than one micron. The purpose of the present letter 
is to analyze whether the manufacturing of such crystals 
is possible within presently existing technologies.

Several techniques have been proposed to produce LALP CU.
At the very beginning  \cite{Kaplin1980a,Baryshevsky1980}, 
it was suggested to use ultrasonic waves to bend the 
crystal. However this idea appeared to be too challenging
from the technical point of view and, therefore, it is 
still waiting for its experimental implementation.
The attenuation of the acoustic wave is inversely 
proportional to its wavelength. i.e. the shorter the 
undulator period, the stronger is the attenuation.
Therefore, using ultrasound in the case of the SASP CU 
would be even more difficult or even impossible. 

A few other technologies utilize the idea of imposing periodic 
stresses on the surface of the crystal sample.
It can be accomplished by
making regularly spaced grooves on the crystal surface
either by a diamond blade \cite{Bellucci2002c,Guidi2005,Bagli2014a}
or by means of laser-ablation \cite{Balling2009}.

Alternatively, the periodic stress can be imposed by 
deposing strips of a different material on the surface
of the crystal. It was initially proposed to use crystalline 
materials with similar but slightly different lattice 
constants \cite{Avakian2002}. Later, it was found more
practical to depose
Si$_3$N$_4$ layers onto the surface of a silicon crystal \cite{Guidi2005}.
Deposition takes place at high temperature. The stress appears after
cooling the crystal to the room temperature
due to different coefficients of thermal expansion of the crystal 
and the deposed material.

Recently, inducing a sub-surface stress of a silicon crystal by 
implantation of He$^{+}$ ions has been studied \cite{Bellucci2015}.

Manufacturing a SASP CU with the surface stress technology 
is hardly possible for the following reasons. 
First, the strain that is produced by the surface stress
decreases fast with the distance from the surface. Therefore, the 
crystal dimension in one of the two transverse directions 
has to be of the order of the undulator period 
$\lambda_\mathrm{u}$ \cite{Kostyuk2007},
which is in the sub-micron range
for moderate beam energies $E \lesssim 1$ GeV due to (\ref{small_lambdau}).  
Preparing such a thin crystal and deposing
sub-micron sized strips on its surface, not to mention making regularly spaced groves, 
are highly problematic. Second, the bending amplitude varies strongly
across the crystal. 
Only the most central part of it having the width of $\sim \lambda_\mathrm{u}/(2 \pi)$
has nearly constant bending amplitude \cite{Kostyuk2007}.  Only this part
should be exposed to
the beam. This means that the size of the beam spot in the corresponding transverse
direction has to be in the range of tens of nanometers and the crystal
has to be placed with the corresponding accuracy. Moreover, as in any other channeling
experiment, the beam divergence 
at the entrance to the crystal should not exceed the critical 
(Lindhard's) angle, which is typically in the range of a few hundreds microradians
or smaller. Therefore, the transverse emittance of the beam should not exceed
several nm$\cdot$mrad. None of the existing electron or positron 
accelerators has such high beam quality \cite{PDG2016}.

Fortunately, there is one more method of crystal bending which is free from 
the above flaws. Growing crystals with varying chemical composition \cite{Mikkelsen2000} 
creates strain inside the crystal volume rather then on its surface.
Therefore, there is no such severe restrictions on the crystal size as in the previous case.
The bending amplitude does not vary across the crystal.
Hence the size of the beam spot can be in the range of
hundreds of microns or even 
a few millimeters in both transverse 
directions. This allows for using  beams of moderate 
quality from presently available accelerators.

The most mature of such technologies is using the method of molecular 
beam epitaxy for growing  Si$_{1-\chi}$Ge$_\chi$ crystals 
with periodically varying germanium fraction $\chi$.
Such heterostructures were intensively studied for the purposes
of the semiconductor industry. There is rather extensive practical 
experience of creating Si$_{1-\chi}$Ge$_\chi$
strained layer of various thickness
from $10${ \AA} to microns or even tens or hundreds of microns. 
Their properties are rather 
well known (see, for instance, \cite{Jain1990,Jain2001,Paul2004} 
and references therein). Deflection of the beam by Si$_{1-\chi}$Ge$_\chi$
strained layer crystals has been studied experimentally, see e.g. 
\cite{Breese1997,Breese1997b}. A Si$_{1-\chi}$Ge$_\chi$ LALP CU 
is being used in ongoing experiments at Mainz Microtron 
\cite{Backe2011,Backe2013,Backe2013a}.
Therefore, the further discussion will be focused on this 
crystalline material. The idea of \cite{Mikkelsen2000} 
will be studied in greater details in order to show 
that it can be used to produce a SASP CU.

\begin{figure}[tbh]
\includegraphics[width=0.95\linewidth]{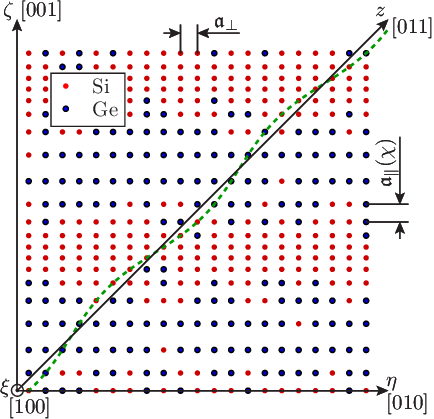}
\caption{A strained layers superlattice of gradually 
varying chemical composition.
A higher fraction $\chi$ of germanium 
results in a larger longitudinal lattice constant
$\mathfrak{a}_{\parallel}(\chi)$. Due to the periodic 
variation of $\chi$, the crystal axis $[011]$ (the dashed line) 
is periodically bent. So is the plane $(01\bar{1})$,
which is perpendicular to the plane of the image and contains 
the axis $[011]$. For illustration purposes, the variation
of the lattice constant is strongly exaggerated, while 
the bending period is shown much shorter
that it should be in the reality.
To simplify the image, only one atom per cubic cell is
shown.
\label{superlattice}}
\end{figure}

Let us consider a Si$_{1-\chi}$Ge$_\chi$ single crystal 
that was grown in the direction [001] with varying germanium 
content $\chi$, see Fig. \ref{superlattice}. 
The coordinate axes $\xi$, $\eta$ and $\zeta$
are chosen to coincide with the crystallographic axes [100],
[010] and [001], respectively. This means that the germanium fraction
$\chi$ depends on $\zeta$ but does not depend on $\xi$ and 
$\eta$.

Let us take an element of the crystal volume containing $\mathcal{N}$
layers of elementary cells in each direction. The volume has to be large 
enough $\mathcal{N} \gg 1$ so that the elasticity theory can be applied.
On the other hand, it should not be too large so that the variation 
of Germanium concentration $\chi$ could
be neglected  within the volume element.

\begin{figure}[tbh]
\includegraphics[width=0.95\linewidth]{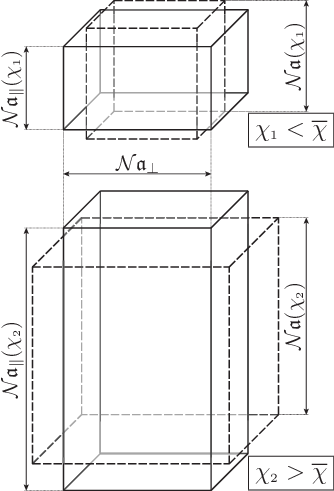}
\caption{Two pieces of a a Si$_{1-\chi}$Ge$_\chi$ crystal having $\mathcal{N}$
layers of elementary crystal cells in each direction. Being relaxed, each piece would 
have a cubic shape (shown by dashed lines). 
Due to different concentrations of germanium, 
$\chi_{\scriptscriptstyle 1}$ and $\chi_{\scriptscriptstyle 2}$, the sizes
of the cubes, $\mathcal{N} \mathfrak{a}(\chi_{\scriptscriptstyle 1})$ and 
$\mathcal{N} \mathfrak{a}(\chi_{\scriptscriptstyle 2})$, would be different.
In a single crystal, both pieces are forced two have the same transverse 
size $\mathcal{N} \mathfrak{a}_{\perp}$. Therefore, they are deformed as
shown by solid lines. The transverse deformation induces a longitudinal 
one. The piece with the smaller than average germanium fraction 
$\chi_{\scriptscriptstyle 1}$ is stretched in the transverse directions,
$\mathfrak{a}_{\perp} > \mathfrak{a}(\chi_{\scriptscriptstyle 1})$,
and contracted in the longitudinal direction, 
$\mathfrak{a}_{\parallel}(\chi_{\scriptscriptstyle 1}) < \mathfrak{a}(\chi_{\scriptscriptstyle 1})$.
In contrast, the piece with the larger than average germanium fraction 
$\chi_{\scriptscriptstyle 2}$ is squeezed in the transverse directions, 
$\mathfrak{a}_{\perp} < \mathfrak{a}(\chi_{\scriptscriptstyle 2})$, and
elongated in the  longitudinal direction 
$\mathfrak{a}_{\parallel}(\chi_{\scriptscriptstyle 2}) > \mathfrak{a}(\chi_{\scriptscriptstyle 2})$.
For illustration purposes, the deformation of the crystal pieces is exaggerated.
\label{cube}}
\end{figure}

Being relaxed, the volume would have a cubic shape with the edge
length $\mathcal{N} \mathfrak{a}(\chi)$ (see Fig. \ref{cube}). 
The 'native' lattice constant 
at given $\chi$, $\mathfrak{a}(\chi)$, 
can be found from a linear interpolation (Vegard's law \cite{Vegard1921})\footnote{A 
small deviation \cite{Dismukes1964} from Vegard's law is neglected in the present analysis.} 
between the lattice
constants of pure silicon $\mathfrak{a}_\mathrm{Si}$ 
and germanium $\mathfrak{a}_\mathrm{Ge}$:
\begin{equation}
\mathfrak{a}(\chi) =  (1-\chi) \mathfrak{a}_\mathrm{Si} + \chi \mathfrak{a}_\mathrm{Ge} .
\label{achi}
\end{equation}

If no defects are present in the crystal, the transverse positions $(\xi,\eta)$
of the atoms in each crystal layer have to 
coincide with those of other layers containing different fraction
of germanium. For this reason, the actual transverse dimension of the volume 
element $\mathcal{N} \mathfrak{a}_{\perp}$
is, generally speaking, different from $\mathcal{N} \mathfrak{a}(\chi)$ 
and does not depend on $\chi$.
Therefore, the volume element is deformed. 

The transverse deformation induces a longitudinal one.
Hence, the longitudinal size of the volume element,
$\mathcal{N} \mathfrak{a}_{\parallel}(\chi)$, is
also, generally speaking, 
different from $\mathcal{N} \mathfrak{a}(\chi)$ (see Fig. \ref{cube}). In contrast 
to $\mathfrak{a}_{\perp}$, the longitudinal size of the volume
element does depend on $\chi$.

The diagonal elements of the strain tensor describing the deformation are
\begin{eqnarray}
\epsilon_{\xi \xi} = \epsilon_{\eta \eta} 
& = & \frac{\mathfrak{a}_{\perp} - \mathfrak{a}(\chi)}{\mathfrak{a}(\chi)} 
\ ,  \label{strainxi} \\
\epsilon_{\zeta \zeta} & = & 
\frac{\mathfrak{a}_{\parallel}(\chi) - \mathfrak{a}(\chi)}{\mathfrak{a}(\chi)}
\label{strainzeta}
\end{eqnarray}
No shear deformation is present.\footnote{In fact, shear deformation is present at the 
edges of (001) planes of the crystal, but it decreases fast with 
the distance from the edge. It is assumed that the sample is sufficiently large
in the directions $\xi$ and $\eta$  so that the shear deformation
can be safely neglected in the bulk of the crystal.}
Therefore non-diagonal elements of the 
strain tensor are zero.

Only three elements of the stiffness tensor $c$ are 
independent in the case of crystals with the cubic symmetry 
(see e.g. \cite{Landau1999}).  
Only two of them, 
$C_{11} = c_{\xi \xi \xi \xi} = c_{\eta \eta \eta \eta} 
= c_{\zeta \zeta \zeta \zeta}
$ and $C_{12} = c_{\xi \xi \eta \eta} = c_{\xi \xi \zeta \zeta} =
 c_{\eta \eta \xi \xi} = c_{\eta \eta \zeta \zeta} =
= c_{\zeta \zeta \xi \xi} = c_{\zeta \zeta \eta \eta}$
are relevant to the present analysis. Hence, Hook's law
for the deformation (\ref{strainxi}) and (\ref{strainzeta})
has the following form
\begin{eqnarray}
\sigma_{\xi \xi} = \sigma_{\eta \eta}  & = &  ( C_{11}(\chi)  + C_{12}(\chi) ) 
\frac{\mathfrak{a}_{\perp} - \mathfrak{a}(\chi)}{\mathfrak{a}(\chi)} \nonumber \\
& & 
+ C_{12}(\chi) \frac{\mathfrak{a}_{\parallel}(\chi) - \mathfrak{a}(\chi)}{\mathfrak{a}(\chi)} \ ,
\label{sigma11} \\
 \sigma_{\zeta \zeta} & = &  2 C_{12}(\chi)   
 \frac{\mathfrak{a}_{\perp} - \mathfrak{a}(\chi)}{\mathfrak{a}(\chi)} \nonumber \\
& & 
+ C_{11}(\chi) \frac{\mathfrak{a}_{\parallel}(\chi) - \mathfrak{a}(\chi)}{\mathfrak{a}(\chi)} \ .
\label{sigma33}
\end{eqnarray}
The stress tensor $\sigma$ has to satisfy the following conditions 
to ensure the mechanical equilibrium of the crystal
\begin{eqnarray}
\overline{\sigma_{\xi \xi}} = \overline{\sigma_{\eta \eta}}  & = & 0 \label{sigma11z} \\
 \sigma_{\zeta \zeta} & = &  0 .  \label{sigma33z}
\end{eqnarray}
The overline stands for averaging over $\zeta$, e.g.
\begin{equation}
\overline{\sigma_{\xi \xi}} = \frac{1}{L_{[001]}}
\int_{0}^{L_{[001]}} \sigma_{\xi \xi} d \zeta \ .
\end{equation}
Here $L_{[001]}$ is the crystal dimension along the crystallographic
axis [001].

The elements of stiffness matrix, $C_{11}(\chi)$ and $C_{12}(\chi)$, depend on the  
germanium concentration $\chi$. Because the mechanical properties 
of silicon and germanium are rather similar, this dependence can be neglected.
The average values $\overline{C_{11}}$ and $\overline{C_{12}}$
will be used in the following.

From (\ref{sigma33}) and (\ref{sigma33z}) one obtains 
\begin{equation}
\mathfrak{a}_{\parallel} (\chi) = \mathfrak{a} (\chi)
- 2 \frac{\overline{C_{12}}}{\overline{C_{11}}}
(\mathfrak{a}_{\perp} - \mathfrak{a} (\chi))
\label{aparchi}
\end{equation}

Substituting the last expression into 
(\ref{sigma33}) and (\ref{sigma33z}) yields
\begin{equation}
\overline{
\left(
\frac{\mathfrak{a}_{\perp} - \mathfrak{a}(\chi)}{\mathfrak{a}(\chi)}
\right)
} = 0. \label{aperpbar}
\end{equation}
The mismatch between the lattice constants of silicon and germanium 
is small as well. Therefore, the variation of $\mathfrak{a}(\chi)$
in the denominator can be also neglected.
Hence, one obtains from (\ref{aperpbar}) and (\ref{achi})
\begin{equation}
\mathfrak{a}_{\perp} = \overline{\mathfrak{a}} = 
(1- \overline{\chi}) \mathfrak{a}_\mathrm{Si} + 
\overline{\chi} \mathfrak{a}_\mathrm{Ge} .
\label{aperp}
\end{equation}

Substituting the last expression into (\ref{aparchi}) one obtains
\begin{equation}
\mathfrak{a}_{\parallel} (\chi) = 
\overline{\mathfrak{a}} +
\tilde{\chi}
\left(
1 + 2 \frac{\overline{C_{12}}}{\overline{C_{11}}}
\right)
\Delta
\label{aparchitild}
\end{equation}
with
\begin{equation}
\tilde{\chi} = \chi - \overline{\chi}
\end{equation}
and
\begin{equation}
\Delta = \mathfrak{a}_\mathrm{Ge} - \mathfrak{a}_\mathrm{Si}.
\end{equation}

\begin{figure}[tbh]
\includegraphics[width=0.95\linewidth]{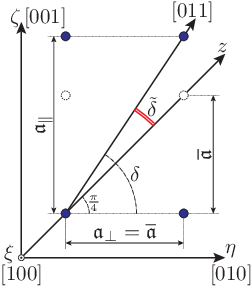}
\caption{
The illustration to the definition of the angles $\delta$ and
$\tilde{\delta}$. 
Due to varying germanium concentration, the 
size $\mathfrak{a}_{\parallel}$
of the crystal cell along the axis $\zeta$ 
is, in general case, different from its average value 
$\overline{\mathfrak{a}}$.
Therefore, the angle $\delta$ between the crystallographic axes 
$[010]$ and $[011]$ varies around its average value $\pi/4$.
The angle between the axis $[011]$ and its average direction
$z$ is $\tilde{\delta}$ (\ref{tildedelta}).
\label{angles}}
\end{figure}

The angle $\delta$ between the axes [010] and [011] (cf. Fig. \ref{angles})
satisfies the following equality
\begin{equation}
\tan \delta = \frac{\mathfrak{a}_{\parallel} (\chi)}{\mathfrak{a}_{\perp}} 
=
1
+
\tilde{\chi}
\left(
1 + 2 \frac{\overline{C_{12}}}{\overline{C_{11}}}
\right)
\frac{\Delta}{\overline{\mathfrak{a}}} .
\label{cotdelta}
\end{equation}
Let us direct the coordinate axis $z$ along the 
average direction of the bent crystallographic axis [011], 
as it is shown in Fig. \ref{angles}.
The axis $z$ makes the angle $\pi/4$ with the axis $\eta$. 
Therefore, the angle
between the axis $z$ and the bent 
 crystallographic axis [011] is 
\begin{equation}
\tilde{\delta} = \delta - \pi/4 .
\label{tildedelta}
\end{equation}
Substituting the last expression into (\ref{cotdelta}) and 
neglecting the terms of the order of $\tilde{\delta}^{2}$
or higher yield
\begin{equation}
\tilde{\delta} = 
\tilde{\chi}
\left(
\frac{1}{2} +  
\frac{\overline{C_{12}}}{\overline{C_{11}}}
\right)
\frac{\Delta}{\overline{\mathfrak{a}}} .
\end{equation}

Let the function $y(z)$ describe the shape of the 
crystallographic axis [011]. Then
\begin{equation}
\frac{d y}{d z} = - \tan \tilde{\delta} \approx - \tilde{\delta}.
\end{equation}
Therefore, 
\begin{equation}
y(z) = y(0) - \left(
\frac{1}{2} +  
\frac{\overline{C_{12}}}{\overline{C_{11}}}
\right)
\frac{\Delta}{\overline{\mathfrak{a}}} 
\int_{0}^{z} \tilde{\chi}(\mathsf{z}) d \mathsf{z} .
\end{equation}

In particular, if $\chi$ varies harmonically with the 
amplitude $\tilde{\chi}_{0}$, 
\begin{equation}
\chi(z) = \overline{\chi} + \tilde{\chi}(z) =
\overline{\chi} + \tilde{\chi}_{0} 
\sin \left( 2 \pi \frac{z}{\lambda_\mathrm{u}}
\right ) \, ,
\end{equation}
the shape of the bent axis is also harmonic:
\begin{equation}
y(z) = y(0) + a_\mathrm{u}  
\cos \left( 2 \pi \frac{z}{\lambda_\mathrm{u}}
\right ) \, ,
\end{equation}
with the bending amplitude 
\begin{equation}
a_\mathrm{u} =  
\left(
\frac{1}{2} +  
\frac{\overline{C_{12}}}{\overline{C_{11}}}
\right) \,
\frac{\Delta}{\overline{\mathfrak{a}}} \,
\frac{\lambda_\mathrm{u}}{2 \pi} \, \tilde{\chi}_{0} \, .
\label{au}
\end{equation}

Let us rewrite the last expression in the following 
form\footnote{The term $\overline{C_{12}}/\overline{C_{11}}$
in the parentheses of Eq.(\ref{chitilde0}) appeared due to the 
longitudinal deformation 
that is induced by the transverse one (see Fig. \ref{cube} and
the two paragraphs after
Eq. (\ref{achi})). The induced deformation was not taken 
into account in \cite{Krause2001b}. It was assumed there that 
$\mathfrak{a}_{\parallel}(\chi)=\mathfrak{a}(\chi)$.
For this reason, the 
necessary germanium content was overestimated by a factor
of about $1.8$.}
\begin{equation}
\tilde{\chi}_{0} =
2 \pi
\frac{\overline{\mathfrak{a}}}{\Delta} \,
\left(
\frac{1}{2} +  
\frac{\overline{C_{12}}}{\overline{C_{11}}}
\right)^{-1} \, 
\frac{a_\mathrm{u}}{\lambda_\mathrm{u}} \, .
\label{chitilde0}
\end{equation}
The average quantities in the right hand side 
depend on $\overline{\chi}$.  

Let us first consider
a small-amplitude variation of 
$\chi(z)$ between $0$ (pure silicon) and 
$2 \tilde{\chi}_{0}$: 
\begin{equation}
 \overline{\chi} = \tilde{\chi}_{0} \ll 1 \, .
\end{equation}
In this case, 
$\overline{\chi}=0$ can be substituted into the 
right hand side of equation (\ref{chitilde0}), 
i.e. the parameters of pure silicon can 
be used instead of the average quantities.

\begin{table}[tb]
\caption{Parameters of silicon and germanium:
the lattice constant $\mathfrak{a}$ and stiffness
coefficients $C_{11}$ and $C_{12}$ \cite{Dismukes1964,Hull1999,Claeys2007}.}
\label{properties}
\begin{center}
\begin{tabular}{|c|c|c|c|}
\hline
   & $\mathfrak{a}$ (\AA) & $C_{11}$ (GPa) & $C_{12}$ (GPa) \\
\hline
Si &     5.431      & 165.6   &  63.9 \\
Ge &     5.658      & 126.0   &  44.0  \\
\hline
\end{tabular}
\end{center}
\end{table}

Using the numerical values from Table \ref{properties}
one obtains
\begin{equation}
\tilde{\chi}_{0} = 170
\frac{a_\mathrm{u}}{\lambda_\mathrm{u}} \, .
\label{chitilde0Si}
\end{equation}

The opposite limit, 
\begin{equation}
 1 - \overline{\chi} = \tilde{\chi}_{0} \ll 1 \, ,
\end{equation}
corresponds to 
a small-amplitude variation of 
$\chi(z)$ between $1 - 2 \tilde{\chi}_{0}$  and 
$1$ (pure germanium). Substituting the parameters from the last
line of Table \ref{properties} into (\ref{chitilde0}) yields
\begin{equation}
\tilde{\chi}_{0} = 184
\frac{a_\mathrm{u}}{\lambda_\mathrm{u}} \, .
\label{chitilde0Ge}
\end{equation}

Interpolating linearly between two extreme cases 
(\ref{chitilde0Si}) and (\ref{chitilde0Ge}) one obtains the
formula 
\begin{equation}
\tilde{\chi}_{0} = \left [ 170 (1 - \overline{\chi})
+ 184 \overline{\chi} \right ]
\frac{a_\mathrm{u}}{\lambda_\mathrm{u}} \, .
\label{chitilde0any}
\end{equation}
that is valid for any $\overline{\chi}$, 
$0 \leq \overline{\chi} \leq 1$. 

The bending period $\lambda_\mathrm{u}$ and the amplitude $a_\mathrm{u}$
refer to the axial channel $[011]$. 
They also valid for the bent planar channel 
$(01\bar{1})$, provided that the beam is directed into this channel
at a small angle to the axis  $[011]$.

It should be stressed that the period of the variation of $\chi$ along 
the direction of crystal growth $[001]$ is smaller
than $\lambda_\mathrm{u}$:
\begin{equation}
\lambda_{[001]} = \frac{\lambda_\mathrm{u}}{\sqrt{2}}.
\label{lambda001}
\end{equation}
Therefore, 
\begin{equation}
\chi(\zeta) = 
\overline{\chi}  -
\tilde{\chi}_{0}
\sin \left( 2 \pi \sqrt{2} \frac{\zeta}{\lambda_\mathrm{u}}
\right )  .
\label{chizeta}
\end{equation}

Formulas  (\ref{chitilde0any}) and (\ref{chizeta}) contain all the necessary
information for calculation of the parameters that are needed for manufacturing 
a strained layer crystalline undulator with desired 
$\lambda_\mathrm{u}$ and $a_\mathrm{u}$. 

Still, the amplitude 
$a_\mathrm{u}$ cannot be arbitrary large. There is
a maximum theoretically possible value $a_\mathrm{th}$
of $a_\mathrm{u}$ at given $\lambda_\mathrm{u}$.
The theoretical limit is set by $\tilde{\chi}_{0}=0.5$
which is the maximum possible value of $\tilde{\chi}_{0}$ corresponding
to $\chi$ varying between $0$ and $1$. Substituting $\tilde{\chi}_{0}=\overline{\chi}=0.5$ 
into (\ref{chitilde0any}), one obtains 
\begin{equation}
a_\mathrm{th} = 2.82 \cdot 10^{-3} \lambda_\mathrm{u}.
\label{ath}
\end{equation}

However, the theoretical limit usually cannot be reached in practice. It is well 
known that there exist a critical thickness $h_\mathrm{c}$ of a strained 
crystal layer in crystalline heterostructures. 
If the layer thickness exceeds $h_\mathrm{c}$, dislocations
appear in the crystal lattice that relax the strain. 

Let us first consider a strained Si$_{1 - \chi}$Ge$_{\chi}$ 
layer grown on a pure silicon substrate.
The larger the germanium  fraction in the layer,  
the smaller its critical thickness.
To put it differently, there exists a critical value of $\chi$
for a given layer thickness $h$ such that the layer 
of thickness $h$ is relaxed with dislocations if its germanium 
fraction $\chi$ is larger than critical. 

The critical layer thickness for stable crystalline heterostructures were 
studied in \cite{Merwe1963b,Matthews1974,Jain1992}. The obtained results
differ only slightly. The formula based on the approach of 
J.~W.~Matthews and A.~E.~Blakeslee
\cite{Matthews1974} 
will be used in the present analysis. The critical germanium 
fraction for a stable Si$_{1-\chi}$Ge$_\chi$
layer of thickness $h$ grown on a pure silicon substrate
is \cite{Houghton1989}
\begin{equation}
\chi_\mathrm{s} = \frac{5.5 \mbox{ \AA}}{h} \ln \frac{h}{1 \mbox{ \AA}} \ 
\label{chistable}
\end{equation}
(the subscript `s' stands for `stable').

To apply formula (\ref{chistable}) to the crystalline undulator, the following 
two points have to be taken into accounts. First, the transverse lattice 
constant of the crystalline undulator is not equal to that of the pure
silicon. It corresponds to the average germanium fraction 
$\overline{\chi}$. Therefore the deviation $\tilde{\chi}$ from 
the average concentration has to be compared to $\chi_\mathrm{s}$ 
instead of $\chi$.
Second, formula (\ref{chistable}) assumes
a constant germanium concentration in the epitaxial layer, while 
$\tilde{\chi}$ varies 
between $0$ and $\tilde{\chi}_{0}$ (or between
$-\tilde{\chi}_{0}$ and $0$) within a half-period. Therefore, the 
average value over the half period 
\begin{equation}
\langle \tilde{\chi} \rangle = \frac{2}{\lambda_{[001]}} 
\int_0^{\lambda_{[001]}/2} \tilde{\chi}_{0}
\sin \left( 2 \pi  \frac{\zeta}{\lambda_{[001]}}
\right ) d \zeta = \frac{2}{\pi} \tilde{\chi}_{0}
\label{chtav}
\end{equation}
will be substituted into the left hand side of (\ref{chistable}).
The length of the half-period $\lambda_{[001]}/2$
has to be substituted for $h$. Finally, taking into account (\ref{chitilde0any})
and (\ref{lambda001}) leads to
the following expression for the maximum bending amplitude of a stable 
crystalline undulator
\begin{equation}
a_\mathrm{s} =( 0.14 \mbox{ \AA} )
\ \ln \left( \frac{\lambda_\mathrm{u}}{2 \sqrt{2} \mbox{ \AA}} \right).
\label{as}
\end{equation}
Note that the influence of $\overline{\chi}$ on the value  of $a_\mathrm{s}$ is
comparable to the rounding error. The value $\overline{\chi}=0.5$ was used in the above
calculation.

%Because the bending amplitude is proportional to the germanium concentration,
%this means that there exists a maximum amplitude of the undulator bending
%for a given undulator period. 

It was found in experiment \cite{Bean1984a}
that dislocation-free epilayers of
much higher thickness and germanium content could be grown
than it had been predicted 
by Matthews-Blakeslee formula (\ref{chistable}).
The reason for it was the kinetic barrier that had to
be overcome before a dislocation could be formed.
Therefore, the thickness of a metastable epitaxial layer 
can exceed the critical value for the stable layer by an order
of magnitude or more. In fact, the Matthews-Blakeslee limit reveals itself
only after annealing the specimen for about $30$ min at 
$750$--$900\,^{\circ} \mathrm{C}$ \cite{Houghton1989}.

Several models describing the critical thickness 
(or, equivalently, the critical germanium content)
of a metastable strained layer have been proposed
\cite{People1985,Dodson1987,Houghton1991}.
The formula of People and Bean\footnote{The original formula
\cite{People1986} was written in terms of misfit $f$ related to
the germanium fraction $\chi$ as $f=0.0418 \chi$. Therefore it had a 
different numerical coefficient in the numerator.} 
\cite{People1985,People1986},
\begin{equation}
\chi_\mathrm{m}^{2} = \frac{10.9 \mbox{ \AA}}{h}
\ \ln \left( \frac{h}{4 \mbox{ \AA}} \right),
\label{PeopleBean}
\end{equation}
will be used in the following. It agrees well with
the experimental data \cite{Bean1984a} obtained for 
a Si$_{1 - \chi}$Ge$_{\chi}$
layer grown by molecular beam epitaxy at 
$550\,^{\circ} \mathrm{C}$.

Repeating the steps that led to formula (\ref{as})
one obtains the maximum amplitude of a metastable
crystalline undulator
\begin{equation}
a_\mathrm{m} =
\sqrt{
( 2.4 \cdot 10^{-3} \mbox{ \AA} ) \lambda_\mathrm{u}
\ \ln \left( \frac{\lambda_\mathrm{u}}{8 \sqrt{2} \mbox{ \AA}} \right)
} \ .
\label{am}
\end{equation}

\begin{figure}[tbh]
\includegraphics[width=0.95\linewidth]{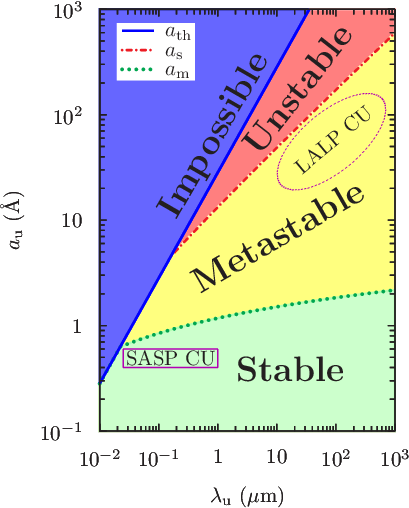}
\caption{The stability diagram of strained layer 
crystalline undulator based on graded 
Si$_{1 - \chi}$Ge$_{\chi}$ composition vs. 
the bending period $\lambda_\mathrm{u}$ of the 
plane $(01\bar{1})$ and the bending amplitude 
$a_\mathrm{u}$ (see the text for details).
The crystalline undulator with a small amplitude and
and a short period (SASP CU) is predicted to be stable,
while that with a large amplitude and
and a long period (LALP CU) is metastable.
\label{domains}}
\end{figure}

Equations (\ref{ath}), (\ref{as}) and (\ref{am}) are summarized in
Fig. \ref{domains}.

The region $a_\mathrm{u}>a_\mathrm{th}$ of the diagram is marked as `Impossible'. 
Undulators with such parameters would require the maximum germanium fraction to
exceed $100${\%}. 

The region 
$a_\mathrm{m} < a_\mathrm{u} < a_\mathrm{th}$ is called `Unstable'.
The crystals with the corresponding germanium fraction can be grown, but 
its crystalline structure will be distorted by misfit dislocations.
As a result, the desired shape of the crystal channel would not be 
obtained. Therefore, such crystals cannot be used as crystalline undulators.

The `Metastable' region corresponds to 
$a_\mathrm{s} < a_\mathrm{u} < a_\mathrm{m},\ a_\mathrm{th}$. Such crystals
can be grown and the desired channel shape can be obtained. But the quality
of such crystalline undulators may degrade with time. The factors that 
facilitate the nucleation of dislocations, e.g. heat and ionizing 
radiation, may accelerate the aging of the strained layer crystal. 

Finally, the undulators satisfying $a_\mathrm{u} < a_\mathrm{s},\ a_\mathrm{th}$ 
are characterized as `Stable'. They are not expected to degrade with time.
Moderately high temperatures, not very close to the melting point, are
not expected to damage such undulators. Just in opposite, annealing may even 
improve their quality.

As is seen from the figure, SASP CU, having 
$a_\mathrm{u} = 0.4$--$0.6${ \AA} \cite{Kostyuk2012}, 
is located in the 'Stable' region of the diagram.
In contrast, LALP CU, that requires $a_\mathrm{u} \gg d = 1.92${ \AA},
can be only metastable.

It has to be stressed that the above statement as well as the whole
preceding consideration are valid if the superlattice is macroscopically 
relaxed, i.e. if there is no external stress acting on it. 
This is possible if a Si$_{1 - \chi}$Ge$_{\chi}$ 
crystal with germanium fraction equal to its average value 
in the superlattice $\chi=\overline{\chi}$ is used as a 
substrate. In this case, the transverse lattice constant 
$\mathfrak{a}_{\perp}$ of the relaxed supelattice and the 
substrate are equal. Due to this fact, the substrate does not 
excert stress on the supelattice.
Therefore, the \textit{total} thickness of the 
superlattice, i.e. the number of undulator periods, is not limited 
by the instability of the epitaxial layer against misfit dislocations.

In contrast, using a pure silicon crystal as a substrate makes even 
SASP CU metastable and limits the total thickness of the 
superlattice. For example, the optimal 
bending parameters for 855 MeV projectiles, 
$\lambda_\mathrm{u}=400$~nm and $a_\mathrm{u} = 0.4$~{\AA}  for electron and 
$\lambda_\mathrm{u}=600$~nm and $a_\mathrm{u} = 0.6$~{\AA}  for positron 
\cite{Kostyuk2012}, can be obtained at $\overline{\chi}=0.017$.
Using equation (\ref{PeopleBean}), one finds that the total thickness
of a metastable superlattice has to be smaller than $44$ $\mu$m.
In this case,
the undulator length (e.g. the superlattice size along the [011] axis)
cannot exceed  $62$ $\mu$m.

If one restricts oneself to a thin metastable epitaxial layer,
another problem arises if a pure silicon substrate is used.

\begin{figure}[tbh]
\includegraphics[width=0.95\linewidth]{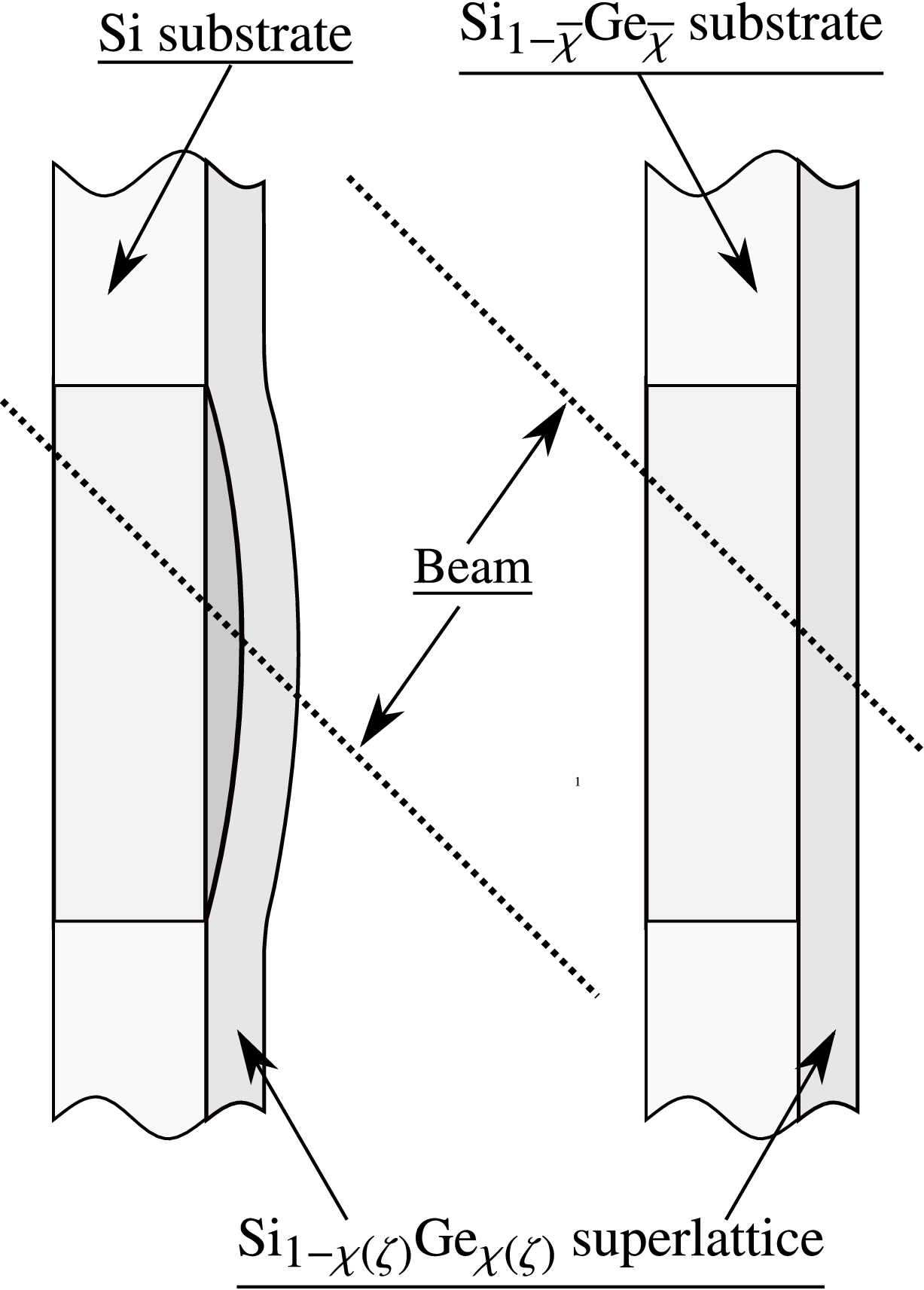}
\caption{A cross section of the window in the substrate
crystal
that lets the beam to cross the Si$_{1 - \chi}$Ge$_{\chi}$ 
superlattice without crossing the substrate.\newline
\textbf{Left:} Due to size mismatch between the 
window in the pure silicon substrate 
and the relaxed superlattice the latter 
experiences a macroscopic deformation (depicted with a great 
exaggeration for illustration purposes).\newline
\textbf{Right:} No size mismatch and, consequently, no macroscopic deformation
are present if a mixed Si$_{1 - \overline{\chi}}$Ge$_{\overline{\chi}}$ crystal
with the germanium fraction $\overline{\chi}$ equal to the average germanium 
fraction in the heterostructure is used as a substrate.
\label{window}}
\end{figure}

To let the beam to cross the superlattice without traversing the substrate,
a `window' has to be made in the substrate.
Otherwise, additional bremsstrahlung and channeling radiation would be produced 
contributing to the undesirable  background. 
Even if microscopic properties of the superlattice 
film in the `window' are not influenced by the substrate any more, the 
size mismatch between the `window' in the pure silicon substrate and the relaxed
superlattice may course a macroscopic deformation of the thin epitaxial layer
(figure \ref{window}, left). As a result, a variation of the 
channel entrance angle exceeding $130$~mrad at $\overline{\chi}=0.017$ 
may occur.
Manufacturing a heterostructure that would be
stable against misfit dislocations and the macroscopic deformation simultaneously,
if is possible at all, would require a very careful design of the `window' profile and 
detailed calculations.

No size mismatch and, consequently, no macroscopic deformation of the 
supperlattice takes place
if a Si$_{1 - \overline{\chi}}$Ge$_{\overline{\chi}}$ crystal
is used as a substrate (figure \ref{window}, right).
Hence, the Si$_{1 - \overline{\chi}}$Ge$_{\overline{\chi}}$ has a double 
advantage: it does not put any limit on the thickness of the epitaxial layer 
and it does not induce macroscopic deformations.  

To conclude, a crystalline undulator can be produced by growing a 
strained layer crystal of a graded
Si$_{1 - \chi}$Ge$_{\chi}$ composition.  The germanium content $\chi$ has to 
be varied according to Eq. (\ref{chizeta}) with the variation amplitude $\tilde{\chi}_{0}$ 
calculated from Eq. (\ref{chitilde0any}).
The obtained crystalline undulators with 
a small amplitudeand a short period (SASP CU)
and 
with a large amplitude and a long period (LALP CU)
are predicted to be, respectively, stable and metastable against  
misfit dislocations, provided that the strained 
layer crystal is grown on a Si$_{1 - \overline{\chi}}$Ge$_{\overline{\chi}}$ substrate 
with germanium content equal to its average value $\overline{\chi}$ 
in the superlattice. In addition, using 
the Si$_{1 - \overline{\chi}}$Ge$_{\overline{\chi}}$ prevents macroscopic deformations 
of the epitaxial layer.

%\begin{acknowledgments}
 \section*{Acknowledgement}
I am grateful to Hartmut Backe for an encouraging discussion.
%\end{acknowledgments}

\end{document}